\documentclass[extra]{gji}
\usepackage{timet,color,bm,amsmath,amssymb, graphicx, multirow}
\usepackage[urlcolor=blue,citecolor=black,linkcolor=black, hidelinks]{hyperref}
\title[DAS-N2N: Machine learning DAS denoising without clean data]{DAS-N2N: Machine learning Distributed Acoustic Sensing (DAS) signal denoising without clean data}
\author[S. Lapins et al.]
  {S. Lapins$^1$,
  A. Butcher$^1$,
  J.-M. Kendall$^2$,
  T.S. Hudson$^2$,
  A.L. Stork$^3$,
  M.J. Werner$^1$, \and J. Gunning$^1$ and A.M. Brisbourne$^4$ \\
  $^1$ School of Earth Sciences, University of Bristol, Bristol, UK. Email: \href{mailto:sacha.lapins@bristol.ac.uk}{\textcolor{blue}{\underline{sacha.lapins@bristol.ac.uk}}}\\
  $^2$ Department of Earth Sciences, University of Oxford, Oxford, UK\\
  $^3$ Silixa Ltd, Bristol, UK\\
  $^4$ British Antarctic Survey, Cambridge, UK
  }

\date{Accepted 2023 November 15. Received 2023 October 31; in original form 2023 April 18}
\pagerange{\pageref{firstpage}--\pageref{lastpage}}
\volume{xxx}
\pubyear{xxxx}

\begin{document}

\label{firstpage}

\maketitle

\begin{summary}
 This article presents a weakly supervised machine learning method, which we call DAS-N2N, for suppressing strong random noise in distributed acoustic sensing (DAS) recordings. DAS-N2N requires no manually produced labels (i.e., pre-determined examples of clean event signals or sections of noise) for training and aims to map random noise processes to a chosen summary statistic, such as the distribution mean, median or mode, whilst retaining the true underlying signal. This is achieved by splicing (joining together) two fibres hosted within a single optical cable, recording two noisy copies of the same underlying signal corrupted by different independent realizations of random observational noise. A deep learning model can then be trained using only these two noisy copies of the data to produce a near fully-denoised copy. Once the model is trained, only noisy data from a single fibre is required. Using a dataset from a DAS array deployed on the surface of the Rutford Ice Stream in Antarctica, we demonstrate that DAS-N2N greatly suppresses incoherent noise and enhances the signal-to-noise ratios (SNR) of natural microseismic icequake events. We further show that this approach is inherently more efficient and effective than standard stop/pass band and white noise (e.g., Wiener) filtering routines, as well as a comparable self-supervised learning method based on masking individual DAS channels. Our preferred model for this task is lightweight, processing 30 seconds of data recorded at a sampling frequency of 1000 Hz over 985 channels (approx. 1 km of fiber) in $<$1 s. Due to the high noise levels in DAS recordings, efficient data-driven denoising methods, such as DAS-N2N, will prove essential to time-critical DAS earthquake detection, particularly in the case of microseismic monitoring.
\end{summary}


\section{Introduction}
\label{sec:intro}

Distributed Acoustic Sensing (DAS) is a novel form of seismic monitoring, measuring changes in strain acting along a buried or encased fibre-optic cable through reflectometry. In recent years, DAS has seen a growing range of applications, including passive and active experiments to detect seismic events, monitor urban and anthropogenic activity, image the subsurface, and monitor changes in material and ambient properties \citep[e.g.,][]{Dou_Lindsey_Wagner_Daley_Freifeld_Robertson_Peterson_Ulrich_Martin_Ajo-Franklin_2017, Ajo-Franklin_Dou_Lindsey_Monga_Tracy_Robertson_Rodriguez_Tribaldos_Ulrich_Freifeld_Daley_et_al._2019, Lindsey_Dawe_Ajo-Franklin_2019, Hudson_Baird_Kendall_Kufner_Brisbourne_Smith_Butcher_Chalari_Clarke_2021, Nayak_Ajo-Franklin_Tribaldos_Dobson_Cheng_Dong_Robertson_Monga_Rotermund_Mellors_et_al._2021, Jousset_Currenti_Schwarz_Chalari_Tilmann_Reinsch_Zuccarello_Privitera_Krawczyk_2022, Kennett_2022, Zhou_Butcher_Brisbourne_Kufner_Kendall_Stork_2022, van-den-Ende_Ferrari_Sladen_Richard_2023}. A DAS interrogator unit sends short, finite-duration light pulses along an optical fibre and measures the phase of Rayleigh backscattering caused by small density variations and defects in the fibre \citep{Parker_Shatalin_Farhadiroushan_2014, Hartog_2017, Lindsey_Rademacher_Ajo-Franklin_2020}. The backscattered light from a given section of fibre returns to the interrogator with a predictable two-way travel-time: this allows any changes in the light’s phase or intensity between successive pulses (e.g., from disturbances as a result of incoming seismic waves) to be mapped to specific sections along the fibre within some known precision, known as the ‘gauge length’ \citep[e.g.,][]{Dean_Cuny_Hartog_2017, Hartog_2017, Lindsey_Rademacher_Ajo-Franklin_2020}. In this manner, the entire fibre-optic cable acts as a series of distributed seismic sensors, sampling changes to the strain field acting on the fibre at regularly spaced intervals along its length \citep[typically shorter than the gauge length;][]{Parker_Shatalin_Farhadiroushan_2014, Hartog_2017}, the locations of which are often referred to as ‘channels’.

Optical fibres have many attractive properties for seismic monitoring. They are flexible, durable, highly sensitive to vibrations and changes in strain field \citep{Hartog_2017}, and can extend many kilometres from the DAS interrogator unit and power source \citep[e.g.,][]{Parker_Shatalin_Farhadiroushan_2014, Ajo-Franklin_Dou_Lindsey_Monga_Tracy_Robertson_Rodriguez_Tribaldos_Ulrich_Freifeld_Daley_et_al._2019, Lindsey_Dawe_Ajo-Franklin_2019, Shinohara_Yamada_Akuhara_Mochizuki_Sakai_2022}. As such, they are well-suited to seismic monitoring in harsh or remote environments, such as volcanoes \citep{Jousset_Currenti_Schwarz_Chalari_Tilmann_Reinsch_Zuccarello_Privitera_Krawczyk_2022}, regions with extreme climate \citep[e.g., glacial settings;][]{Walter_Graff_Lindner_Paitz_Kopfli_Chmiel_Fichtner_2020, Hudson_Baird_Kendall_Kufner_Brisbourne_Smith_Butcher_Chalari_Clarke_2021, Zhou_Butcher_Brisbourne_Kufner_Kendall_Stork_2022} and beneath oceans \citep{Lindsey_Dawe_Ajo-Franklin_2019, Shinohara_Yamada_Akuhara_Mochizuki_Sakai_2022}. Furthermore, the vast networks of existing telecommunications fibre present the opportunity to heavily augment the coverage of existing seismic networks, both on local and global scales \citep{Ajo-Franklin_Dou_Lindsey_Monga_Tracy_Robertson_Rodriguez_Tribaldos_Ulrich_Freifeld_Daley_et_al._2019, Nayak_Ajo-Franklin_Tribaldos_Dobson_Cheng_Dong_Robertson_Monga_Rotermund_Mellors_et_al._2021, Kennett_2022, Shinohara_Yamada_Akuhara_Mochizuki_Sakai_2022}.

However, despite these advantageous properties, optical fibres are also highly sensitive to temperature \citep{Hartog_2017}, local disturbances from the interrogator \citep{Lindsey_Rademacher_Ajo-Franklin_2020}, ground / coupling conditions \citep{Hartog_Frignet_Mackie_Clark_2014, Ajo-Franklin_Dou_Lindsey_Monga_Tracy_Robertson_Rodriguez_Tribaldos_Ulrich_Freifeld_Daley_et_al._2019}, and properties of the fibre / instrument components used \citep{Isken_Vasyura-Bathke_Dahm_Heimann_2022}, most of which are heterogeneously distributed along the optical cable, leading to greater levels of seemingly random observational noise in DAS recordings when compared to conventional seismometers \citep{Hudson_Baird_Kendall_Kufner_Brisbourne_Smith_Butcher_Chalari_Clarke_2021}. DAS fibres are also only sensitive to along-cable strain, which leads to challenges in recording the full seismic wavefield and relating measurement units to actual ground motion. Lastly, the large data volumes acquired by sampling along long extents of fibre (sometimes on the order of TBs per day) require highly efficient and optimized signal processing methods, especially for real-time monitoring operations and earthquake early-warning systems.

Optical cables are regularly manufactured with multiple fibres for added capacity (e.g., for telecommunication providers). For DAS applications, additional fibres are typically left unused as DAS interrogators often process measurements from a single light pulse and fibre at a time to avoid interference \citep{Parker_Shatalin_Farhadiroushan_2014}. However, the availability of multiple fibres can provide highly useful redundancy for enhancing the signal-to-noise ratio (SNR) of any external signal through application of so-called ‘weakly supervised’ machine learning. By splicing (joining together) two fibres at one end of the cable, the light pulse from the DAS interrogator effectively travels ‘there-and-back’ along the length of the cable, recording two copies of the same underlying seismic signal but with different random measurement noise due to differences in scatterers and photon behaviour between the two fibres. A deep learning model can then be trained using only these two noisy copies of the underlying signal to produce a denoised copy of the data through a method known as “Noise2Noise” \citep[N2N;][]{Lehtinen_Munkberg_Hasselgren_Laine_Karras_Aittala_Aila_2018}, a form of weakly supervised machine learning that exploits the point estimation properties of certain loss functions and does not require clean (i.e., noise-free) target data or manual curation / labelling for training.

In this article, we present the first known application of N2N for suppressing strong random (i.e., incoherent) noise processes in DAS data, which we refer to as DAS-N2N. This approach has previously been used to suppress synthetically generated noise in individual photographic, MRI scan and microscopy images \citep{Lehtinen_Munkberg_Hasselgren_Laine_Karras_Aittala_Aila_2018, Calvarons_2021} but never previously (to our knowledge) to suppress real noise in continuously acquired noisy DAS or seismic data. In Section \ref{sec:background}, we provide an overview of both ‘conventional’ (i.e., non-machine learning) and machine learning approaches for seismic signal noise suppression, including N2N. In Section \ref{sec:data}, we provide details of our example dataset, acquired by a DAS deployment on the surface of the Rutford Ice Stream in Antarctica. In Section \ref{sec:methods}, we describe the theory behind N2N, and the procedure for training and implementing a DAS-N2N model. In Section \ref{sec:results}, we compare icequake data denoised by DAS-N2N against three benchmark methods: conventional Butterworth bandpass filtering, Wiener filtering, and an existing self-supervised deep learning method for denoising DAS data, known as jDAS \citep{van-den-Ende_Lior_Ampuero_Sladen_Ferrari_Richard_2021}, that also requires no clean target data or manual curation during model training. The article ends with a discussion of DAS-N2N model performance, generalisation to other datasets, and some concluding remarks in Sections \ref{sec:discussion} and \ref{sec:conclusions}, respectively. Example code and data for implementing DAS-N2N have been archived and made available by \citet{Lapins_Butcher_Kendall_Hudson_Stork_Brisbourne_2023} (see Data Availability section).

\section{Background}
\label{sec:background}

\subsection{Conventional seismic signal filtering}
\label{subsec:conventional}

Pass and stop band filters, designed to remove certain frequencies from a recorded signal, are a ubiquitous processing step for suppressing unwanted noise in seismic signals. The general aim is to identify a frequency range that contains as much of the desired signal and as a little of the undesired background noise as possible, with all other frequencies removed or suppressed by a chosen filter (e.g., Butterworth, Chebyshev or Gaussian filters). These filters are typically applied by convolution of the recorded signal with a polynomial approximating an idealized filter response \citep[i.e., approximating a uniform and complete response in the pass band with full attenuation in the stop band, which cannot be expressed by a finite order polynomial;][]{Proakis_Manolakis_1996}. Although simple, interpretable and relatively fast for individual seismic traces, such methods have several drawbacks, both for DAS applications and for seismic signals more generally.

For noise suppression, the greatest drawback of conventional pass and stop band filters is the inability, by design, to suppress noise that lies in the same frequency range as the desired signal. For well-deployed geophones and broadband seismometers, random measurement noise is considered to be low and signals of interest are usually in distinct frequency bands from other external coherent noise sources \citep[e.g., ocean microseisms in the 0.1 – 0.5 Hz range;][]{Bromirski_Stephen_Gerstoft_2013, Koper_Burlacu_2015, Lapins_Roman_Rougier_De-Angelis_Cashman_Kendall_2020}. However, environmental, financial, cultural and political factors mean that deploying large numbers of high-cost seismometers in quiet or well-insulated environments is rarely feasible. DAS offers a relatively low-cost, straightforward and densely sampled alternative; however, random measurement noise along the fibre is often observed to be much stronger than that of geophones \citep{Hudson_Baird_Kendall_Kufner_Brisbourne_Smith_Butcher_Chalari_Clarke_2021, du-Toit_Goldswain_Olivier_2022, Isken_Vasyura-Bathke_Dahm_Heimann_2022} and occurs across the entire observed frequency spectrum (see Section \ref{sec:results}). As such, the frequency range of interest is much more contaminated by unwanted noise. Furthermore, for passive monitoring applications, this frequency range must be assumed \textit{a priori}, which typically leads to more conservative (i.e., wider pass band) filtering and greater noise contamination. Choice of filter family and polynomial order is also subject to certain trade-offs, including the degree of amplitude ‘ripples’ in the pass and stop bands, the abruptness of the transition between bands, and susceptibility to detrimental artefacts such as ringing, signal polarity changes and nonlinear phase shifts \citep[e.g.,][]{Proakis_Manolakis_1996, Scherbaum_2001, Havskov_Ottemoller_2010}. Lastly, and importantly, repeated application of a chosen filter over hundreds or thousands of individual DAS channels is computationally costly when required for (near-)real-time processing and monitoring.

Some of the drawbacks outlined above can be mitigated: e.g., through use of adaptive algorithms that adjust filter coefficients or parameters \citep[e.g.,][]{Duncan_Beresford_1994, Jeng_Li_Chen_Chien_2009, Isken_Vasyura-Bathke_Dahm_Heimann_2022}; filtering in both time and space frequency (f-k) domains \citep[e.g.,][]{Duncan_Beresford_1994, Bacon_Simm_Redshaw_2003, Mousa_2019, Hudson_Baird_Kendall_Kufner_Brisbourne_Smith_Butcher_Chalari_Clarke_2021, Isken_Vasyura-Bathke_Dahm_Heimann_2022}; applying a statistical estimation method for identifying additive or incoherent noise \citep[e.g., Wiener filters;][]{Williams_Kendall_Verdon_Clarke_Stork_2020}; or combining multiple methods \citep{Chen_Savvaidis_Fomel_Chen_Saad_Wang_Oboue_Yang_Chen_2023}. However, these approaches are still limited when noise and signal overlap within a given frequency range or by their computational demands, model / method assumptions, or the requirement for manual parameterization (see Section \ref{sec:results}).

\subsection{Deep learning denoising}
\label{subsec:deeplearning}

Across the broader fields of science and engineering, noise suppression (or ‘denoising’) is being increasingly addressed through deep learning methods, with the greatest advancements occurring in the field of image processing. Unlike linear, f-k (time and space frequency) or statistical estimation filters, deep learning models are not restricted by explicit statistical assumptions, response trade-offs (e.g., choice of filter family or order), or manual parameterization (e.g., choosing a pass band or statistical model). Desirably, they have the capacity to ‘learn’ empirical, abstract and nonlinear hierarchical data representations directly from sample data, allowing them to perform effective signal filtering and feature extraction without manual input or prior assumptions on the distribution of the signal or noise. Model implementation is also heavily optimizable through use of GPUs and compression / pruning strategies \citep{Zhu_Gupta_2017}, allowing for rapid signal processing.

\subsubsection{Supervised learning}
\label{subsubsec:supervised}

Initial success in this area was driven by the ‘standard’ fully supervised paradigm, using a large number of noisy/clean signal pairs for model training; i.e., both noisy and noise-free copies of each training sample are available and the model is trained to directly map noisy signals to their noise-free counterparts. This approach is sometimes referred to as 'Noise2Clean' (N2C) in the denoising literature and has been previously applied to seismic signals with apparent success \citep{Zhu_Mousavi_Beroza_2019, Klochikhina_Crawley_Frolov_Chemingui_Martin_2020, Li_Ma_2021, Tibi_Hammond_Brogan_Young_Koper_2021, Yang_Liu_Zhu_Zhao_Beroza_2022}. However, in many applications, it can be difficult or even impossible to acquire sufficient quantities of noise-free or high signal-to-noise recorded signals for robust model training, and thus this approach is limited in its ‘real-world’ applicability. This situation is particularly true in the case of DAS recordings, where the observed data are heavily contaminated by strong random noise processes and simulating seismic wave propagation to generate realistic noise-free signals across a long extent of fibre is computationally intensive and challenging to model. This restricted applicability has led to the wider development of denoising methods that do not require noise-free ground-truth signals, such as weakly supervised \citep{Zhou_2018, van-Engelen_Hoos_2020} or self-supervised \citep{Ericsson_Gouk_Loy_Hospedales_2022} learning methods.

\subsubsection{Weakly supervised learning}
\label{subsubsec:weaklysupervised}

Weakly supervised learning relaxes the requirement for noise-free ‘ground-truth’ target data during training. One pioneering method for weakly supervised denoising, which we base our proposed DAS-N2N methodology on, is known as 'Noise2Noise' \citep[N2N;][]{Lehtinen_Munkberg_Hasselgren_Laine_Karras_Aittala_Aila_2018}, where the aim is for a model to learn to transform noisy images into clean images using only noisy copies of the same image as both input and target training data. A N2N model suppresses random noise by exploiting the point estimation properties of certain loss functions during model training \citep{Lehtinen_Munkberg_Hasselgren_Laine_Karras_Aittala_Aila_2018, Pang_Zheng_Quan_Ji_2021}; e.g., mean squared error (MSE) and mean absolute error (MAE) loss functions are minimized by the mean and median of a set of observations, respectively. Intuitively, as long as the noise in the input and target data are independently and randomly drawn from some (known or unknown) noise distribution, it is impossible for a model to predict the random noise values in the target data from the random noise values in the input data. As such, to minimize its expected loss, the model learns to map noise in the input data to the value of smallest average deviation from the noise in the target data \citep[e.g., the mean, median or mode of the noise distribution;][]{Lehtinen_Munkberg_Hasselgren_Laine_Karras_Aittala_Aila_2018}, according to the chosen loss function. Simultaneously, as long as the underlying clean signal in the input and target data are identical, the model’s expected loss is minimized by learning a direct 1-to-1 mapping between the two (see Section \ref{subsec:theory}).

It has been demonstrated both theoretically and empirically that models trained using only noisy signals in this manner can perform as well as, or even better than, those trained in a fully supervised manner using noisy/clean signal pairs \citep{Lehtinen_Munkberg_Hasselgren_Laine_Karras_Aittala_Aila_2018, Pang_Zheng_Quan_Ji_2021}. For example, it can be shown that the loss minimization problem is effectively the same for fully and weakly supervised learning and a MSE loss function (Section \ref{subsec:theory}).

\begin{figure}
    \centering
    \includegraphics[width=8.5cm]{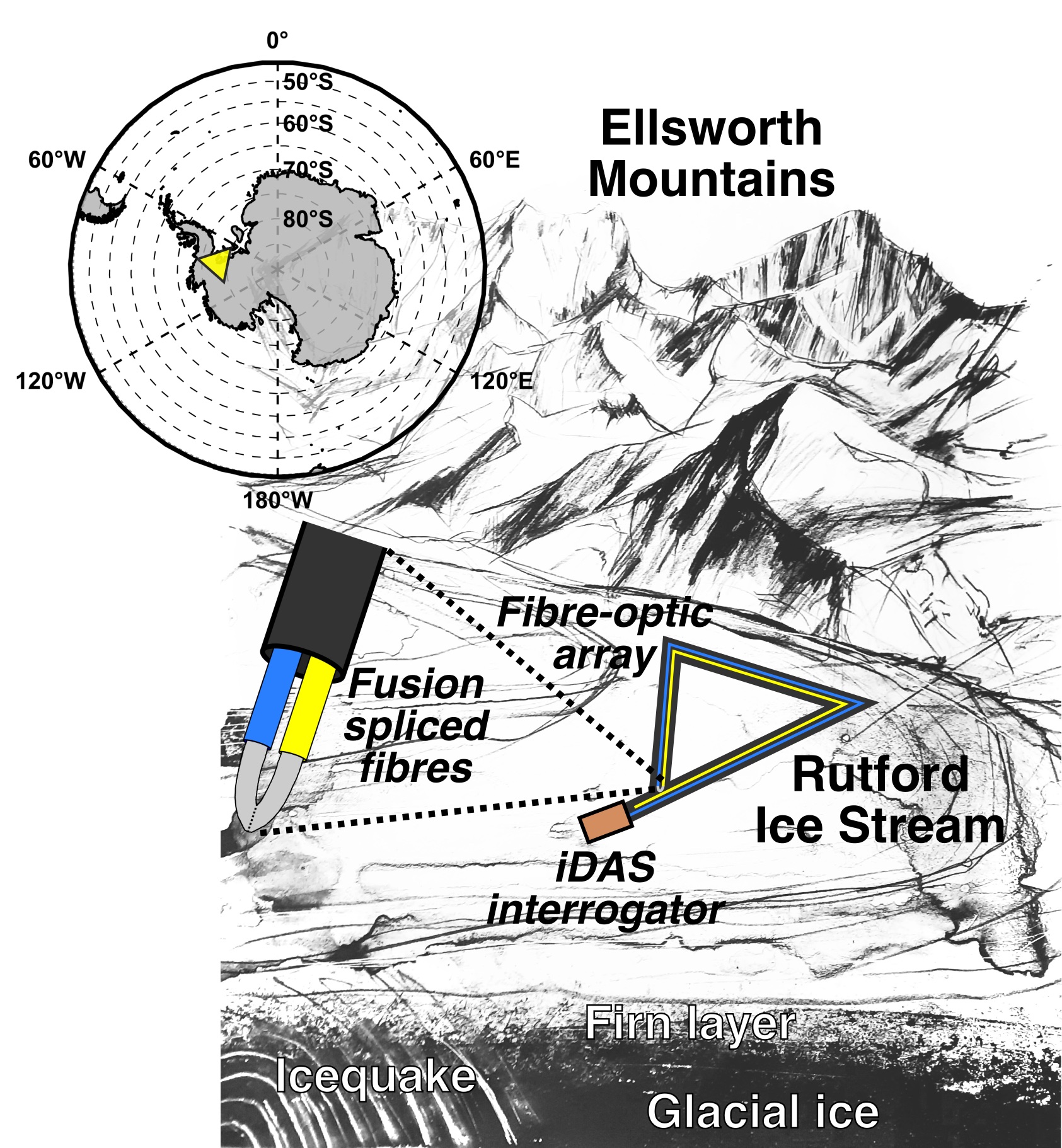}
    \caption{Map and schematic illustration showing the DAS experiment setup. Top left: Map showing geographic location of DAS array (gold triangle) on Rutford Ice Stream, Antarctica. Main: Schematic illustration of the DAS experiment. DAS fibre array was deployed in triangular configuration on the surface of Rutford Ice Stream, with two single-mode fibres hosted within a single cable jacket spliced at cable end. See \citet{Hudson_Baird_Kendall_Kufner_Brisbourne_Smith_Butcher_2021} and Supporting Information in \citet{Hudson_Baird_Kendall_Kufner_Brisbourne_Smith_Butcher_Chalari_Clarke_2021} for further details.}
    \label{fig:fig1}
\end{figure}

\begin{figure}
    \centering
    \includegraphics[width=8.5cm]{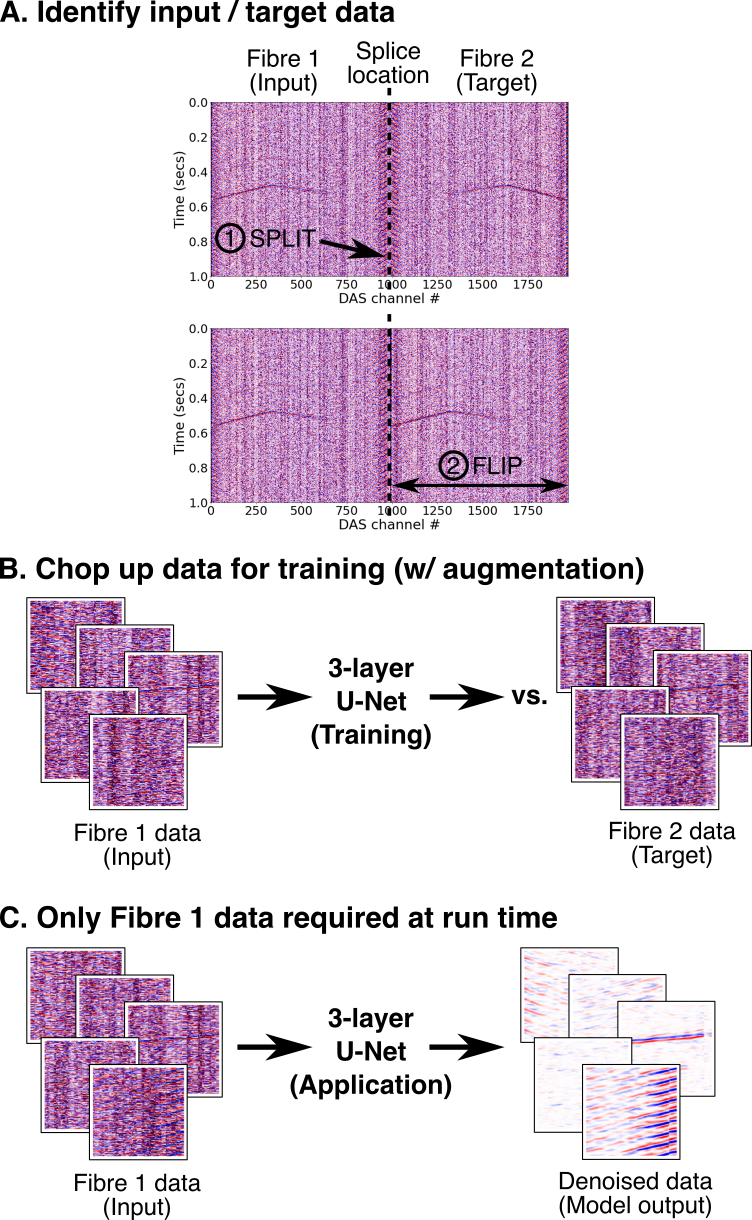}
    \caption{Implementing DAS-N2N. A) Raw data is split into input (Fibre 1) and target (Fibre 2) training data. B) Data is divided into smaller sections (128 samples x 96 channels) for model training, with augmentation (vertical / horizontal flipping) randomly applied to each training sample pair. C) Once the model is trained, only the input data (Fibre 1) is required for denoising.}
    \label{fig:fig2}
\end{figure}

For DAS, the applicability of N2N is motivated by the fact that optical fibres can be spliced so that they effectively double-back on themselves within their cable sleeve (Fig \ref{fig:fig1}), recording two (near-) identical copies of any external seismic source but with different independent realizations of any random noise processes (Fig \ref{fig:fig2}A). Furthermore, when continuous recordings are available, vast training sets of independent noisy signal pairs are readily available for training without the need for any manual labelling, providing a fully automatable approach that can be applied to any DAS deployment.

The one main drawback of N2N is that, in some situations, recording multiple noisy copies of the same underlying signal is not possible; e.g., when analysing previously recorded un-spliced DAS data or using so-called ‘dark’ fibres (existing unused telecommunication fibre networks) that may not always be feasibly spliced. In this case, N2N is not directly applicable, but an extension of this method, based on 're-corrupting' the recorded signal with additional noise \citep[known as ‘Recorrupted-to-Recorrupted’, or R2R;][]{Pang_Zheng_Quan_Ji_2021}, can be applied. With R2R, the additional noisy copies of the signal required for model training are generated (as opposed to recorded) using a secondary noise distribution so that the noise in each new noisy copy is now independently drawn from a new noise distribution. These new noisy copies can then be used to train a model in the same manner as N2N, with similar performance \citep{Pang_Zheng_Quan_Ji_2021}. However, sufficiently corrupting the original observed noise to produce independent realizations from a new noise distribution means one must generally have some prior knowledge of the original noise distribution, which is not always known or may be challenging to model. Due to this added non-trivial requirement, we do not explore this method further in this paper and restrict our focus solely on the N2N approach with recorded noisy signal pairs.

\subsubsection{Self-supervised learning}
\label{subsubsec:self}

An alternative approach that requires no additional noisy or clean target data for training (i.e., an un-spliced fibre can be used) is self-supervised learning. Self-supervised learning is often formulated as learning from 'fill-in-the-gap' problems \citep{Ericsson_Gouk_Loy_Hospedales_2022}, where some section of input data is hidden or masked and the model is tasked with predicting the values of the missing data. When applied to the task of denoising \citep[sometimes known as ‘Noise2Self’ or ‘Noise2Void’;][]{Krull_Buchholz_Jug_2018, Batson_Royer_2019}, the intuition behind such an approach is that, through training, the model will learn to interpolate or predict missing coherent or broad-scale signal features, based on the surrounding data and exposure to many training samples, but will be unable to predict random, incoherent or fine-scale signal features. As with weakly supervised learning, the model minimizes its expected loss by learning to map the latter to the value (or point estimate) of smallest average deviation, according to some loss function (e.g., MSE).

One such self-supervised method, known as jDAS, after the concept of j-invariance \citep{Batson_Royer_2019}, has been previously applied to DAS \citep{van-den-Ende_Lior_Ampuero_Sladen_Ferrari_Richard_2021}. With jDAS, individual DAS channels are dropped during training and the model learns to predict these missing data using data from neighbouring channels; i.e., by effectively learning to interpolate any coherent signal across missing channels. This step of masking and predicting missing data is then repeated for each DAS channel at run-time \citep{van-den-Ende_Lior_Ampuero_Sladen_Ferrari_Richard_2021}. Similarly, \citet{Birnie_Ravasi_Liu_Alkhalifah_2021} apply the same concept by treating dense-array post-stack seismic data as a 2D image and masking rectangular / square sections of the image during training. Alternatively, \citet{Liu_Yue_Zuo_Xu_Fu_Yang_Jiang_2022} divide the data into odd (input data) and even (target data) channel numbers and train a model to map one to the other, effectively amounting to the same task as masking every other channel and predicting the missing data.

Although self-supervised learning has the highly desirable trait of not requiring any additional clean or noisy copies of the data, its effectiveness becomes increasingly limited when noise levels are high \citep{van-den-Ende_Lior_Ampuero_Sladen_Ferrari_Richard_2021} and the number of DAS channels to process is large. The desired signal features of interest must be interpolated by the model, rather than ‘retained’ through a 1-to-1 mapping as in the case of weakly or fully supervised learning, and thus signal quality can suffer as a result of the same point estimation properties (e.g., mapping to average of all possible outcomes) being used to suppress incoherent and fine-scale noise.  Furthermore, by masking and predicting only one channel or a small number of datapoints at a time, self-supervised methods, when formulated as fill-in-the-gap problems, are an order of magnitude slower than standard or weakly-supervised methods and can become prohibitively slow when the signal sample frequency and number of DAS channels to process are high.

\section{Data}
\label{sec:data}

In this article, we demonstrate DAS-N2N using data acquired by a DAS array deployed on the surface of the Rutford Ice Stream in Antarctica (Fig \ref{fig:fig1}). Despite being a low anthropogenic noise environment, strong random noise processes (e.g., optical noise caused by random scattering/coupling of photons and environmental factors) in the raw recorded data from this deployment dominate the signal from microseismic icequake events. The deployment consists of a Silixa iDASv2 interrogator \citep{Parker_Shatalin_Farhadiroushan_2014} and a 1 km cable, with a sample frequency of 1000 Hz, channel spacing of 1 m and gauge length of 10 m \citep[see][for further details]{Hudson_Baird_Kendall_Kufner_Brisbourne_Smith_Butcher_Chalari_Clarke_2021}. Two single-mode optical fibres hosted within the cable jacket were spliced at one end to form a ‘there-and-back’ loop (Fig \ref{fig:fig1}). These data were collected to investigate the suitability of DAS for studying natural microseismicity \citep{Hudson_Baird_Kendall_Kufner_Brisbourne_Smith_Butcher_Chalari_Clarke_2021} and imaging the near-surface ice structure \citep{Zhou_Butcher_Brisbourne_Kufner_Kendall_Stork_2022}.

Over the course of 14 days (2020-01-11 – 2020-01-24), the DAS fibre-optic cable was repeatedly deployed in different horizontal arrangements on the surface of the ice stream, comprising a linear, triangular and ‘hockey stick’ array. The cable was coupled to the ground by placing it in a skidoo track and back-covering with snow. The data presented here were chosen from the time period during which the fibre was deployed in a triangular configuration (2020-01-17 0100 – 0500 UTC), with each linear section of the triangle approximately 330 m in length. Data recorded between 0100 and 0300 UTC were used for model training, with the remaining two hours data used as test data. As the DAS cables were deployed horizontally and P-waves arrive at the surface with near-vertical incidence due to the presence of a low-velocity firn layer, only S-wave phase arrivals are observed by the fibre during the deployment \citep{Hudson_Baird_Kendall_Kufner_Brisbourne_Smith_Butcher_Chalari_Clarke_2021}. The example icequake events presented in Section \ref{sec:results} were detected using a localized Radon-transform-based detection method \citep{Butcher_Hudson_Kendall_Kufner_Brisbourne_Stork_2021}.

\section{Methods}
\label{sec:methods}

\subsection{DAS-N2N thoery}
\label{subsec:theory}

Seismic signals are contaminated by both coherent (i.e., seismic waves generated by some undesired external source) and incoherent (i.e., random) noise. A noisy signal, $\bm{y}$, can be expressed as a sum of independent signal components, such that

\begin{equation} \label{eq:1}
    \bm{y} = \bm{x} + \bm{n},
\end{equation}

\noindent where $\bm{x}$ is a single- or multi-dimensional array representing the underlying ‘clean’ signal from any recordable external seismic source (including external coherent noise sources) and $\bm{n}$ are samples randomly drawn from some noise distribution, with one sample of $\bm{n}$ drawn for each element in $\bm{x}$. The noise distribution is often assumed to be Gaussian, as a result of the Central Limit Theorem, but this is not a requirement for N2N.

When DAS fibres are spliced, a second copy of the underlying signal is near-simultaneously recorded, with

\begin{equation} \label{eq:2}
    \bm{\tilde{y} = \bm{x} + \bm{\tilde{n}}},
\end{equation}

\noindent where $\bm{\tilde{y}}$ is a second noisy copy of clean signal $\bm{x}$, corrupted by random noise samples $\bm{\tilde{n}}$ (drawn independently of noise samples $\bm{n}$). We observe that samples drawn from $\bm{n}$ and $\bm{\tilde{n}}$ need not be locally identically and independently distributed (i.i.d.; see Section \ref{sec:results}).

With DAS-N2N, these two noisy signals, $\bm{y}$ and $\bm{\tilde{y}}$, serve as input and target data, respectively, for training a neural network, $f_\theta$, parameterized by model weights, $\theta$. This neural network is trained to minimise the expected loss between $f_\theta(\bm{y})$ and $\bm{\tilde{y}}$ according to some loss function, $L$. For an MSE loss function (i.e., $L(x,y)=\frac{1}{N}\sum(x-y)^2$), this expected loss can be expressed as

\begin{equation} \label{eq:3}
    \mathop{{}\mathbb{E}} \left\{ L \left[ f_\theta(\bm{y}_i), \bm{\tilde{y}}_i \right] \right\} = \mathop{{}\mathbb{E}} \left\{ \frac{1}{M} \sum_{i=0}^{M} \left[ (\bm{x}_i + \bm{\tilde{n}}_i) - f_\theta(\bm{y}_i) \right] ^2\right\},
\end{equation}

\noindent where $i$ is training sample index and $M$ is the number of training samples in a training batch. Equation \ref{eq:3} can be trivially expanded \citep{Pang_Zheng_Quan_Ji_2021}, such that

\begin{multline} \label{eq:4}
    \mathop{{}\mathbb{E}} \left\{ \frac{1}{M} \sum_{i=0}^{M} \left[ (\bm{x}_i + \bm{\tilde{n}}_i) - f_\theta(\bm{y}_i) \right] ^2\right\} 
    \\= \mathop{{}\mathbb{E}} \left\{ \frac{1}{M} \sum_{i=0}^{M} \left[ \bm{x}_i - f_\theta(\bm{y}_i) \right] ^2\right\} 
    + \mathop{{}\mathbb{E}} \left\{ \frac{2}{M} \sum_{i=0}^{M} \bm{\tilde{n}}_i \bm{x}_i \right\} 
    \\- \mathop{{}\mathbb{E}} \left\{ \frac{2}{M} \sum_{i=0}^{M} \bm{\tilde{n}}_i f_\theta(\bm{y}_i) \right\} 
    + \mathop{{}\mathbb{E}} \left\{ \frac{1}{M} \sum_{i=0}^{M} \bm{\tilde{n}}_{i}^{2} \right\},
\end{multline}

\noindent where the first term, $\mathop{{}\mathbb{E}} \left\{ \frac{1}{M} \sum_{i=0}^{M} \left[ \bm{x}_i - f_\theta(\bm{y}_i) \right] ^2 \right\}$, is equivalent to the expected MSE loss when training using noisy/clean training pairs (i.e., the standard supervised case). As long as $\bm{n}$ and $\bm{\tilde{n}}$ are independent, the remaining expectation terms are constant \citep{Pang_Zheng_Quan_Ji_2021}: the two intermediate terms are equal to zero if signal, noise and model output are all zero-mean (enforced by simple subtraction of recorded signal mean, a near-ubiquitous seismic pre-processing step) and summed over sufficiently large $M$, with the final expectation term equal to the variance of the noise distribution in the target data. As such, the loss minimization task when training a model with DAS-N2N can be expressed as

\begin{equation} \label{eq:5}
    \mathop{{}\mathbb{E}} \left\{ L \left[ f_\theta(\bm{y}_i), \bm{\tilde{y}}_i \right] \right\} = \mathop{{}\mathbb{E}} \left\{ \frac{1}{M} \sum_{i=0}^{M} \left[ \bm{x}_i - f_\theta(\bm{y}_i) \right] ^2\right\} + c,
\end{equation}

\noindent which is equivalent to the standard noisy/clean supervised case, up to a constant, $c$, relating to the variance of the noise. It is for this reason that DAS-N2N can perform as well as a model trained with noisy/clean signal data, with the advantage that all recorded data can be used for model training without any manual curation or the need to ‘generate’ noisy/clean signal pairs.

\subsection{Implementing DAS-N2N}
\label{subsec:implementing}

As mentioned, a DAS-N2N model is trained by using data recorded by one of the spliced fibres as input data, with data recorded by the other spliced fibre as target data (Fig \ref{fig:fig2}A). The only pre-processing steps applied in this work are to remove the signal mean (across all channels) and normalize the data (i.e., divide through by the standard deviation). When training a model using data from longer fibres that have spatially changing or highly non-linear noise processes (e.g., from hanging sections or light decay) channel-wise normalisation will likely be required to ensure the data remain centered around zero and consistently normalised. The raw data were originally stored as 30 s TDMS files \citep[the standard file type for data acquired using a Silixa iDAS interrogator;][]{Parker_Shatalin_Farhadiroushan_2014}, and thus these pre-processing steps are applied to 30 s sections of data at a time.

The input and target data are then split into corresponding 128 x 96 size arrays (no. of time samples x no. of DAS channels, respectively), with a batch size of 24 used for model training (Fig \ref{fig:fig2}B). Training data are augmented by randomly flipping both the input and target data along their vertical (time) and horizontal (channel) axes. The loss between the model-processed input data and the noisy target data is calculated for each batch using an MSE loss function, with model weights updated using the Adam optimization algorithm \citep{Kingma_Ba_2014}. The model was trained for 30 epochs, with learning rate decreasing between epochs from $10^{-3}$ to $10^{-5}$ over the course of model training.

N2N is based on exploiting the point estimation properties of L2 and L1 loss functions \citep{Lehtinen_Munkberg_Hasselgren_Laine_Karras_Aittala_Aila_2018}, and therefore its performance is relatively agnostic to choice of model architecture (i.e., any model with sufficient capacity can be trained to perform N2N denoising). However, certain model architectures and components will have advantageous qualities for denoising, such as hierarchical feature representation (e.g., from convolutional layers) and use of dense / residual connections (e.g., to retain underlying signal as it passes from layer to layer). With this in mind, we choose to implement DAS-N2N using a shallow, 3-layer U-Net \citep[][see \textit{Appendix A: Model Architecture}]{Ronneberger_Fischer_Brox_2015}. By limiting the number of model layers and using skip connections, the underlying signal can be easily retained from layer-to-layer and computational processing time is kept low. The final 3-layer DAS-N2N model has just 47,065 model parameters, processing 30 seconds of recorded data across 985 channels (30,000 x 985 data points) in $<$ 1 s (average processing time over 10 runs using Python 3.7.12, TensorFlow version 2.3.0 \citep{tensorflow2015-whitepaper} and a single NVIDIA GeForce RTX 2080 Ti GPU).

Following training, only the input data (i.e., data from a single fibre) are required for data processing (Fig \ref{fig:fig2}C). The normalization step applied during training is also reversed at this stage (i.e., the model-processed data are multiplied by the original data standard deviation) to recover absolute signal amplitude.

\subsection{jDAS implementation}
\label{subsec:jdas}

For comparison with our proposed DAS-N2N methodology, we implement the self-supervised jDAS approach described by \citet{van-den-Ende_Lior_Ampuero_Sladen_Ferrari_Richard_2021}, using the same model architecture as our DAS-N2N model and applying the same data normalization / augmentation steps (Section \ref{subsec:implementing}), to serve as a benchmark for comparable ‘noisy data only’ machine learning approaches. The data are split into 2048 x 11 data blocks for model training, as proposed by \citet{van-den-Ende_Lior_Ampuero_Sladen_Ferrari_Richard_2021}, with a mask randomly applied to a single DAS channel for each training sample.

Both the DAS-N2N and jDAS models are trained using the same 3-layer U-Net architecture, MSE loss function, Adam optimizer, learning rate schedule and number of epochs for direct comparison of method effectiveness.

\subsection{Conventional bandpass filtering}
\label{subsec:bpfiltering}

For comparison with standard seismic filtering steps, we bandpass filter the raw DAS data between 10 and 100 Hz, based on icequake signal characteristics from \citet{Hudson_Baird_Kendall_Kufner_Brisbourne_Smith_Butcher_Chalari_Clarke_2021}, using a 4th order Butterworth infinite impulse response (IIR) filter. We apply a two-pass filter to remove any nonlinear phase shift, allowing for more direct comparison between methods (Section \ref{sec:results}). Filtering is performed using the open-source ObsPy Python library \citep{Beyreuther_Barsch_Krischer_Megies_Behr_Wassermann_2010, Megies_Beyreuther_Barsch_Krischer_Wassermann_2011, Krischer_Megies_Barsch_Beyreuther_Lecocq_Caudron_Wassermann_2015}, which uses optimized low-level C programming language routines from the popular and widely used SciPy library \citep{Virtanen_Gommers_Oliphant_Haberland_Reddy_Cournapeau_Burovski_Peterson_Weckesser_Bright_et_al._2020}. Butterworth filters have a near-uniform response in the pass band and are thus a popular choice for seismic signal processing as they adequately retain underlying signal amplitude information used for further seismic signal analysis (e.g., earthquake magnitude estimation). This near-uniform response also provides a benchmark for comparing absolute signal amplitudes against DAS-N2N and jDAS processing methods.

\subsection{Wiener filtering}
\label{subsec:wienerfiltering}

Wiener filtering is a classical technique for removing additive white noise and is commonly used to suppress unwanted incoherent noise in seismic data. These filters estimate the power of the underlying signal and additive noise by calculating the mean and variance over localised regions of the data, and optimise the separation of these processes through minimising an MSE loss function. These filters work best when the noise is constant-power ("white") additive noise, such as Gaussian noise, and provide a useful comparison for benchmarking the performance of DAS-N2N for suppressing incoherent noise in raw DAS data.

We apply a Wiener filter to our data with a window size of 7x7, which is the size of the receptive field for the 3-layer U-Net used to implement DAS-N2N and jDAS (i.e., the area of input data that a deep learning model can 'see', given its depth, filter kernel size, etc). Smaller window sizes are less effective at suppressing incoherent noise, and larger window sizes more aggressively suppress underlying signal. Filtering is performed using the widely-used, open-source SciPy library \citep{Virtanen_Gommers_Oliphant_Haberland_Reddy_Cournapeau_Burovski_Peterson_Weckesser_Bright_et_al._2020}.

\section{Results}
\label{sec:results}

\subsection{Denoising example \#1 (in-sample data)}
\label{subsec:example1}

\begin{figure*}
    \centering
    \includegraphics[width=13.3cm]{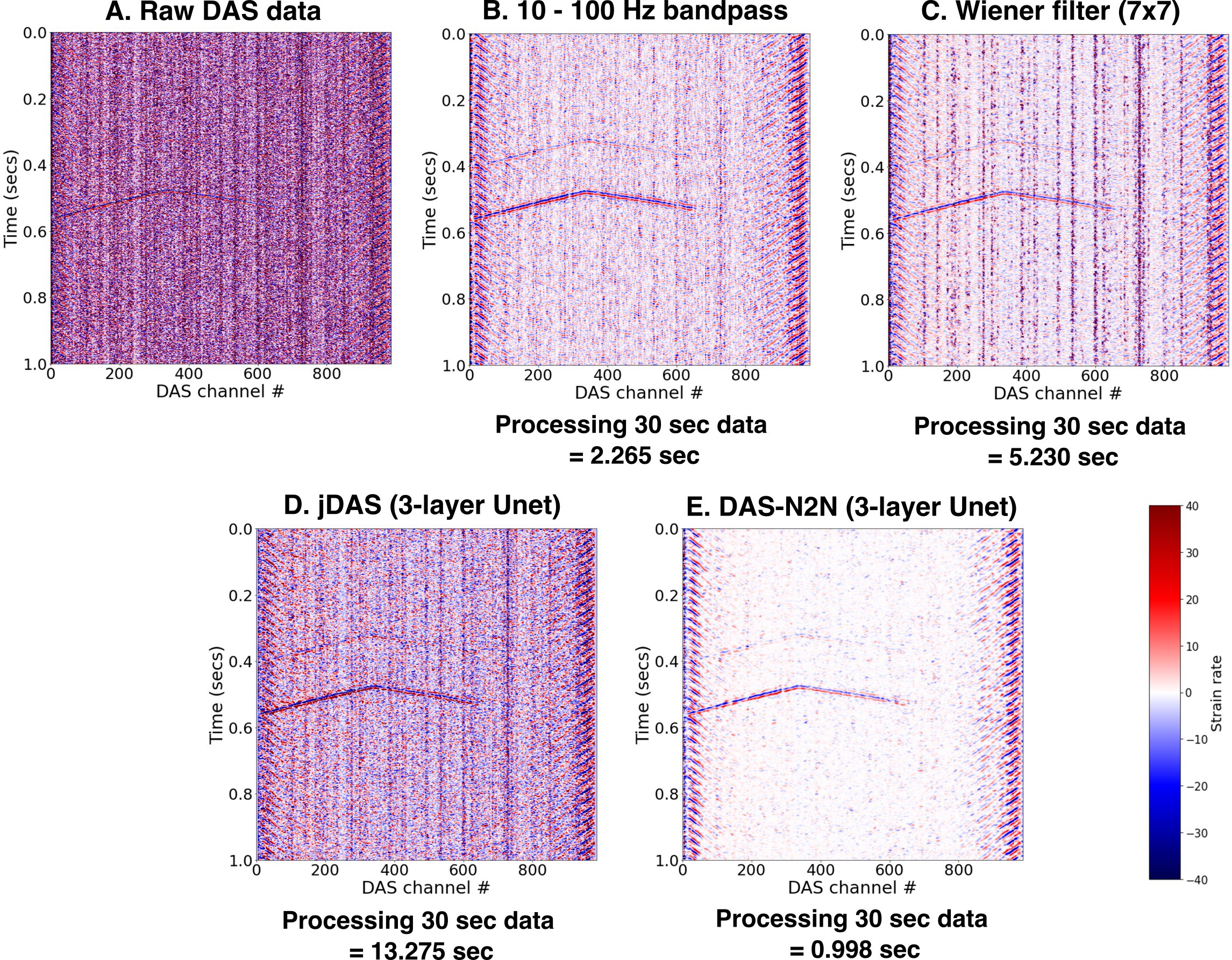}
    \caption{In-sample example of two icequakes (S-wave arrivals only) recorded by DAS deployment (time in seconds after 2020-01-17 01:30:19.232 UTC). A) Raw DAS data. B) Butterworth (2-pass, 4th order) 10 – 100 Hz bandpass filtered DAS data. C) Wiener filtered (7x7 window size) DAS data. D) jDAS filtered DAS data. E) DAS-N2N filtered DAS data. Icequake S-waves arrive at DAS channel 0 at time 0.4 s and 0.55 s, respectively. Strain rate is recorded in units of strain/s (counts).}
    \label{fig:fig3}
\end{figure*}

Figure \ref{fig:fig3} shows the S-wave arrivals from two icequakes recorded by the DAS deployment. These two events occur during the two hours of continuous data (2020-01-17 0100 – 0300 UTC) used for model training and are thus regarded as ‘in-sample’ data (i.e., data the model has ‘seen’ during training).

From Figure \ref{fig:fig3}, it is clear that the raw recorded DAS data are corrupted by very strong random noise (Fig \ref{fig:fig3}A), with the two S-wave arrivals (arriving at approx. 0.4 s and 0.55 s on DAS channel 0, respectively) almost completely indistinguishable from the random background noise. The intensity of this noise varies spatially along the fibre but appears to show some uniformity over small sections (i.e., vertical streaks of high intensity noise are visible over multiple contiguous channels). This suggests that the noise in these data may not be independent across neighbouring channels (a required assumption for jDAS individual channel masking procedure).

Application of bandpass or Wiener filtering (Fig \ref{fig:fig3}B and \ref{fig:fig3}C) clearly improves the signal-to-noise ratio (SNR) of these arrivals, along with that of the surface waves produced by the DAS power generator visible at each end of the fibre (channels 0 – 100 and 850 – 986). However, the higher intensity vertical noise streaks present in the raw data are still present, particularly in the Wiener filtered data. SNR also appears to be improved over the raw data when using either the jDAS (Fig \ref{fig:fig3}D) or DAS-N2N (Fig \ref{fig:fig3}E) models, although the degree of noise suppression clearly differs between methods. Data processed by the jDAS model appears to still be strongly contaminated by random noise, including the same higher intensity noise streaks present in the raw and bandpass/Wiener filtered data, whereas, of all the methods presented, the DAS-N2N model appears to perform the greatest degree of background noise suppression (Fig \ref{fig:fig3}D), without any discernible noise streaks (noisy channels), and is therefore likely to yield the greatest improvement in SNR.

To confirm these observations, we examine estimates of local SNR determined using semblance \citep[a measure of signal similarity across DAS channels;][]{Neidell_Taner_1971}. A moving window of size 19 samples x 13 channels is applied to the data, with channel-wise cross-correlation and a minimum correlation coefficient of 0.7 used to correct for any local moveout within a window. Semblance is then calculated for each moveout-corrected window using the formula

\begin{equation} \label{eq:6}
    S = \frac{ \sum_{i=1}^{N} \left( \sum_{j=1}^{M} x_{ij} \right) ^2 }{ M \sum_{i=1}^{N} \sum_{j=1}^{M} x_{ij}^{2}},
\end{equation}

\noindent where $x_{ij}$ is the moveout-corrected DAS data with time index $i$ and DAS channel $j$. Equation \ref{eq:6} effectively represents the ratio of signal coherency to total signal energy. This value can then be used to estimate local SNR \citep{Bakulin_Silvestrov_Protasov_2022} by

\begin{equation} \label{eq:7}
    \text{SNR}_{\text{local}} = S / (1-S).
\end{equation}

\begin{figure*}
    \centering
    \includegraphics[width=13.3cm]{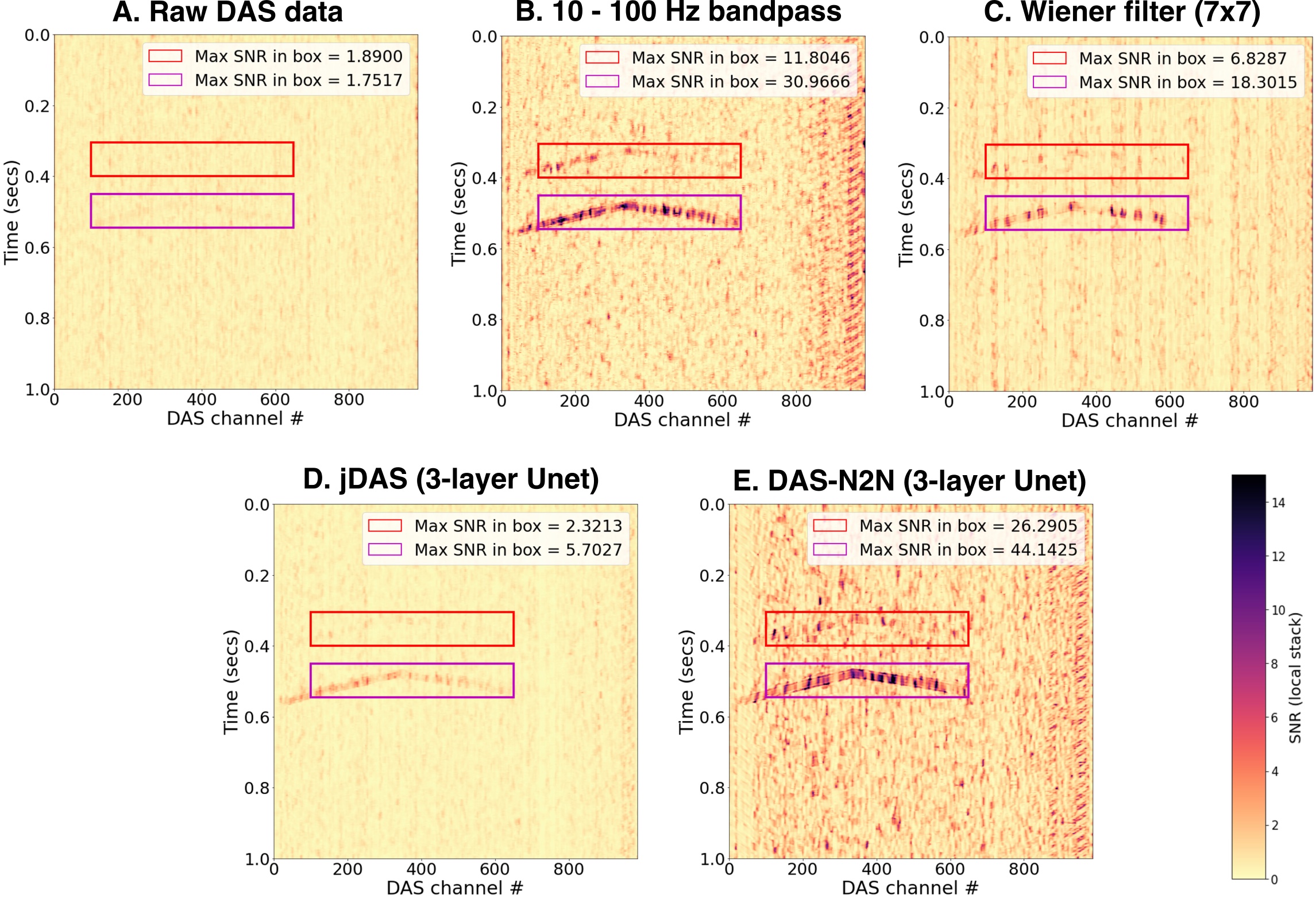}
    \caption{Local signal-to-noise ratio (SNR) estimates for each example in Figure \ref{fig:fig3}. SNR is calculated using semblance (Equation \ref{eq:7}) and a 13-channel x 19-sample 2D moving window}
    \label{fig:fig4}
\end{figure*}

Intuitively, when random noise levels are low, coherent phase arrival signals will be very similar across neighbouring DAS channels, resulting in a high semblance score, $S$, and thus a high estimate of SNR, according to Equation \ref{eq:7}. On the other hand, signals that are corrupted by strong random noise will have lower similarity across neighbouring channels and therefore yield a lower semblance score, $S$, resulting in a lower estimate of SNR. Of all the methods used, DAS-N2N results in the highest SNR for these two S-wave arrivals (Fig \ref{fig:fig4}E). This is true regardless of window size or chosen summary statistic (e.g., maximum, mean, median or quantile) used to compare local SNR for an arrival, and is also true for all events examined across this DAS deployment. The jDAS model (Fig \ref{fig:fig4}D) yields a higher SNR than the raw data (Fig \ref{fig:fig4}A) but fails to suppress background noise as well as conventional bandpass filtering (Fig \ref{fig:fig4}B), Wiener filtering (Fig \ref{fig:fig4}C) and DAS-N2N (Fig \ref{fig:fig4}E).

\begin{figure*}
    \centering
    \includegraphics[width=13.3cm]{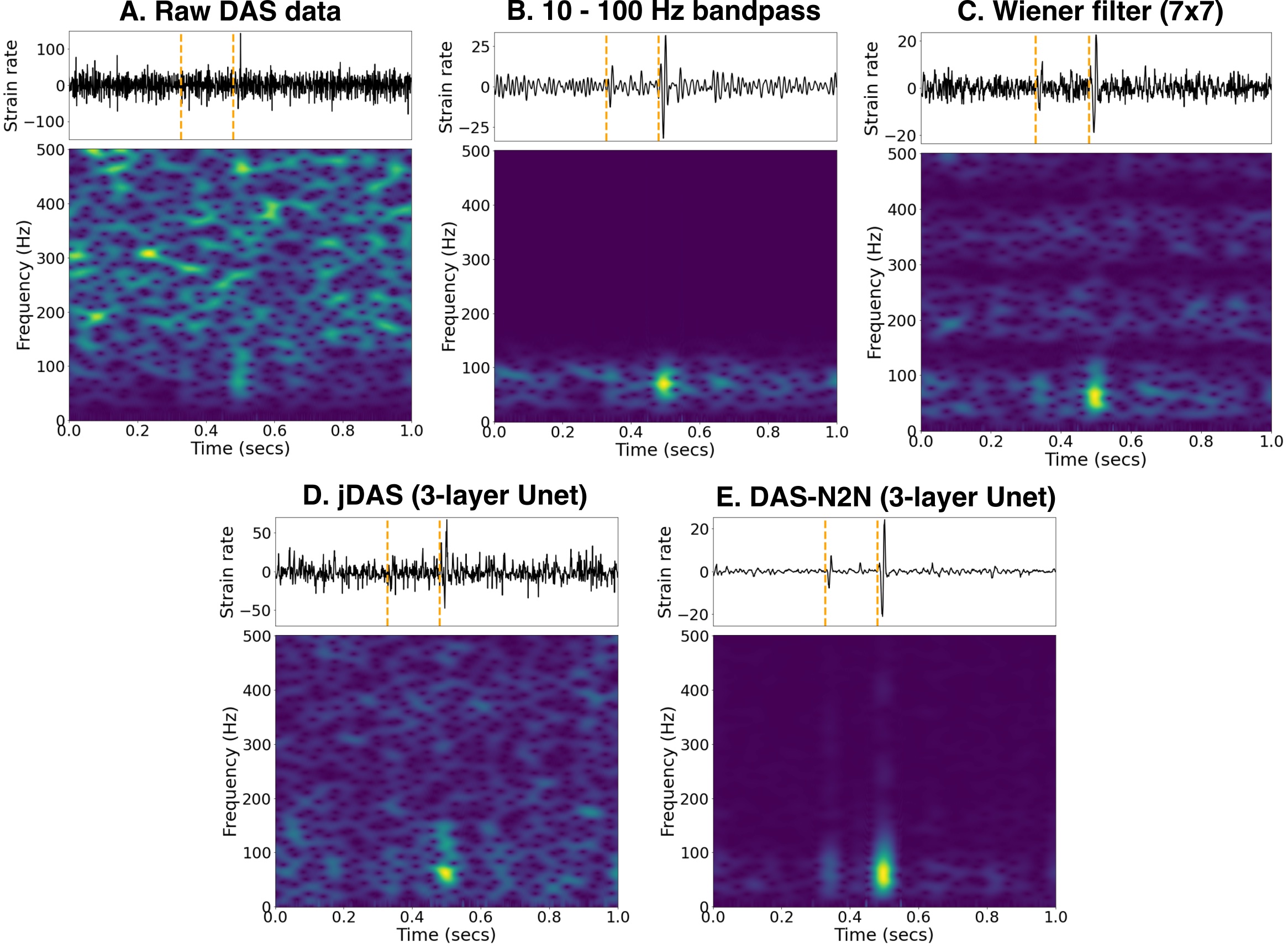}
    \caption{Individual DAS trace (top) and corresponding spectrogram (bottom) for DAS channel 255 in each example in Figure \ref{fig:fig3}. Strain rate is recorded in units of strain/s (counts).}
    \label{fig:fig5}
\end{figure*}

Figure \ref{fig:fig5} shows a single DAS channel trace for each noise suppression method (top of each panel), along with their corresponding time-frequency spectrograms (bottom of each panel). From the spectrogram of the raw data (Fig \ref{fig:fig5}A), the random measurement noise appears to follow a ‘blue noise’ process, with the power or intensity of the noise increasing with frequency and remaining (locally) time-invariant. This observation could be useful for other machine learning DAS denoising methods, such as R2R or generating noisy/clean signal pairs for supervised learning, where the recorded signal must be corrupted to generate new, independent noise samples for model training.

From these individual traces and spectrograms, it is evident that the DAS-N2N model yields the greatest degree of noise suppression, with both S-wave arrivals clearly visible against background noise in both time (Fig \ref{fig:fig5}E top) and time-frequency (Fig \ref{fig:fig5}E bottom) domains. Furthermore, the DAS-N2N spectrogram (Fig \ref{fig:fig5}E bottom) demonstrates that this method goes beyond simple spectral filtering: noise within the 10 – 100 Hz range, which encompasses the dominant frequencies of the two recorded phase arrivals, is also greatly suppressed when compared with bandpass filtering (Fig \ref{fig:fig5}B), and low-amplitude high-frequency signal components ($>$100 Hz) are also retained. It is in this manner that DAS-N2N and other machine learning methods can exceed the performance of conventional stop/pass band filtering.

Although, in relative terms, DAS-N2N signals are stronger (with respect to background noise), absolute signal amplitudes after DAS-N2N processing are weaker than their corresponding bandpass filter benchmark (by a factor of approx. 4/5; see vertical-axis labels on traces in Fig \ref{fig:fig5}). This 4/5 scaling appears to be consistent across all events examined from this deployment \citep{Butcher_Hudson_Kendall_Kufner_Brisbourne_Stork_2021}. This amplitude difference likely relates to signal leakage, where some of the desired underlying signal is suppressed with the noise, and is a common issue with denoising methods based on MAE and MSE loss functions \citep[e.g.,][]{Birnie_Alkhalifah_2022}, and where the raw data are very noisy or large regions of the underlying data are 'empty' (i.e., vast majority of the data contain no events; see Discussion in Section \ref{sec:discussion}). A similar degree of signal leakage occurs with the Wiener filtered data (Fig \ref{fig:fig5}C), which is also optimised using an MSE loss function.

\subsection{Denoising example \#2 (out-of-sample data)}
\label{subsec:example2}

\begin{figure*}
    \centering
    \includegraphics[width=13.3cm]{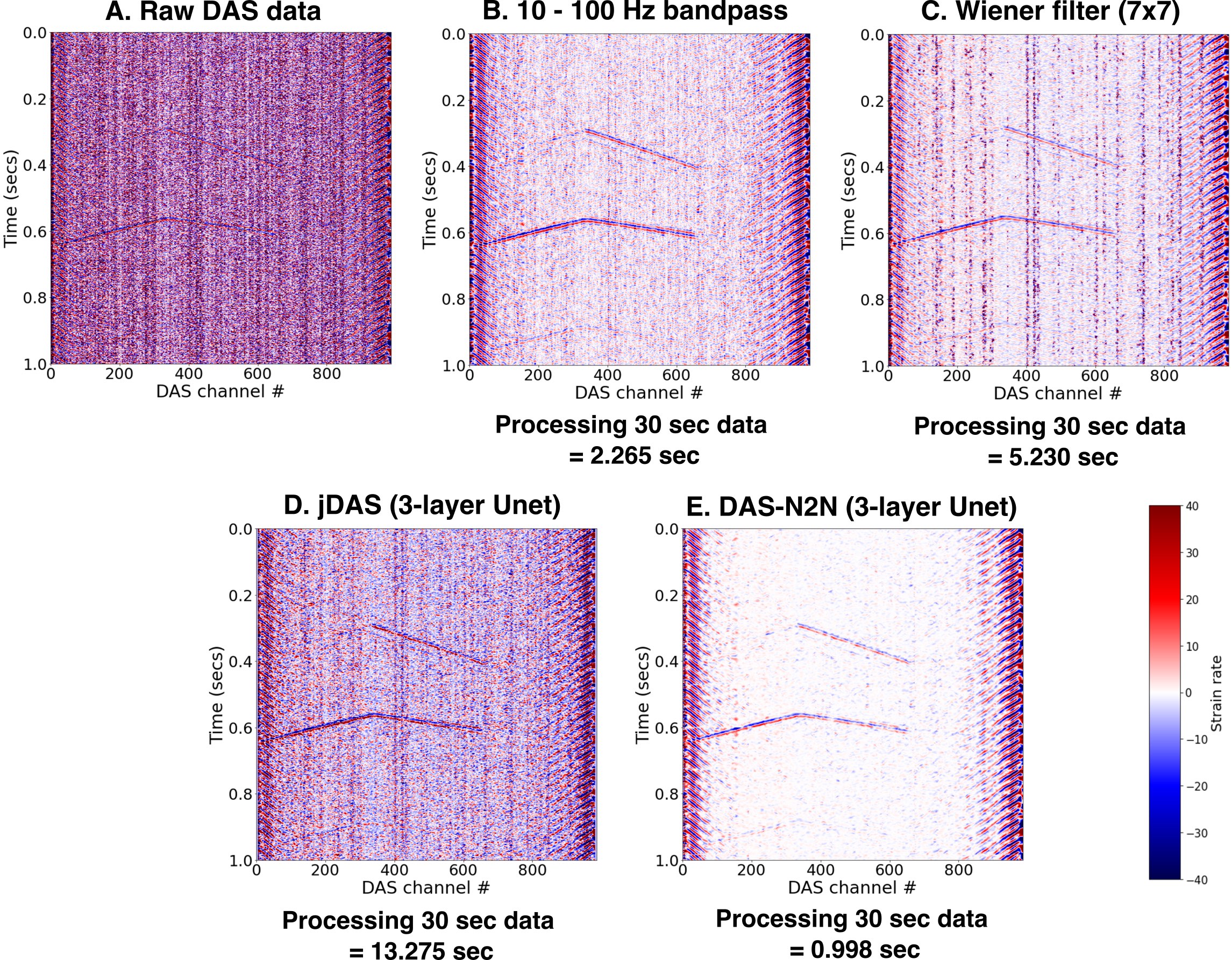}
    \caption{Out-of-sample example of three icequakes (S-wave arrivals only) recorded by DAS deployment (time in seconds after 2020-01-17 04:42:07.903 UTC). A) Raw DAS data. B) Butterworth (2-pass, 4th order) 10 – 100 Hz bandpass filtered DAS data. C)  Wiener filtered (7x7 window size) DAS data. D) jDAS filtered DAS data. E) DAS-N2N filtered DAS data. Icequake S-waves arrive at DAS channel 0 at time 0.4 s, 0.63 s and 0.95 s, respectively. Strain rate is recorded in units of strain/s (counts).}
    \label{fig:fig6}
\end{figure*}

In Figure \ref{fig:fig6}, we present another short section of recorded DAS data from a time period outside of our training set (2020-01-17 04:42:07.903 UTC). This section was chosen to demonstrate the performance of our model on so-called ‘out-of-sample’ data, with three S-wave arrivals from discrete icequake events (arrival times on DAS channel 0 at approx. 0.4 s, 0.63 s and 0.95 s, respectively) observed in the bandpass/Wiener filtered, jDAS denoised and DAS-N2N denoised data (Figs \ref{fig:fig6}B – E).

\begin{figure*}
    \centering
    \includegraphics[width=13.3cm]{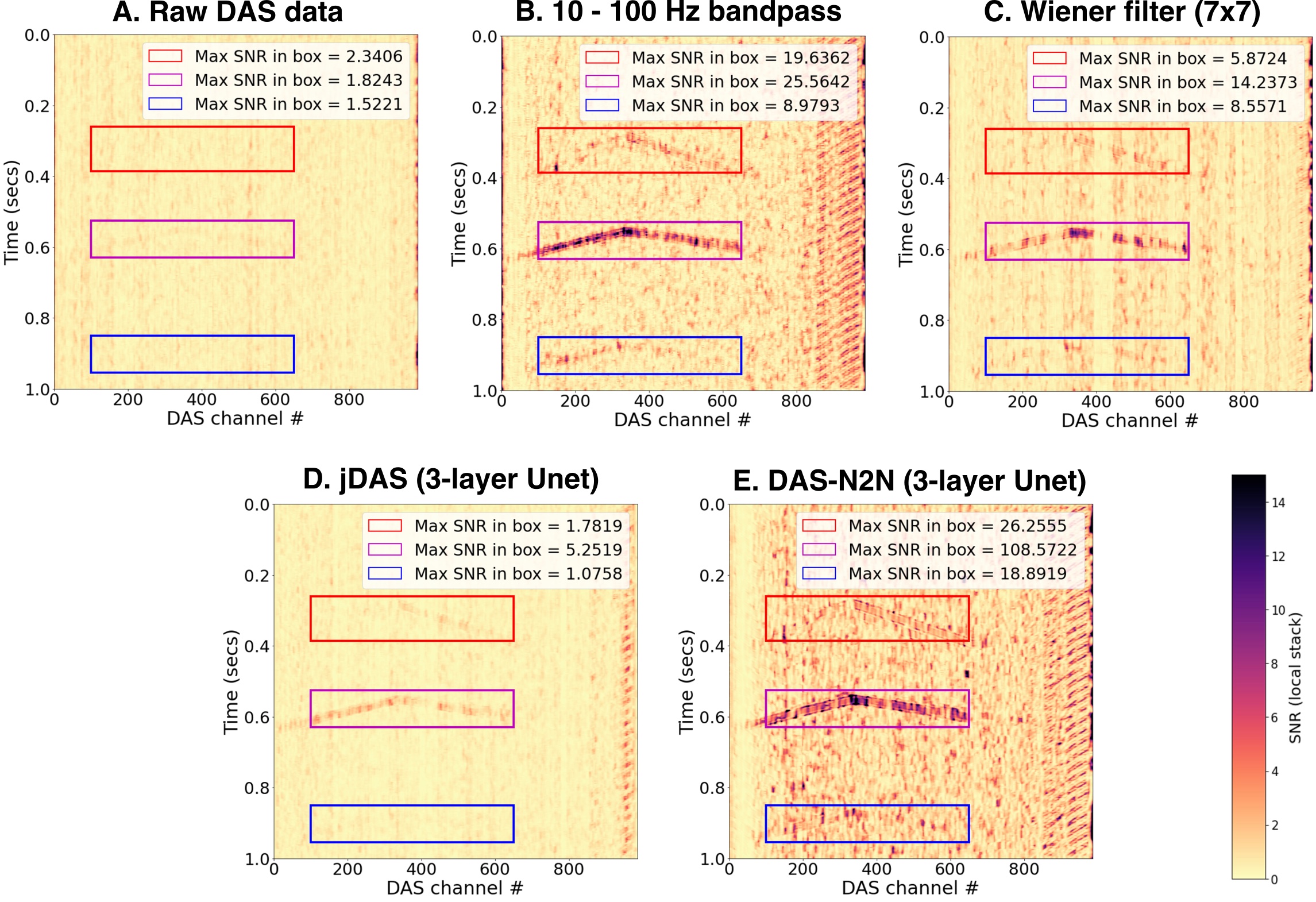}
    \caption{Local SNR estimates for each example in Figure \ref{fig:fig6}. SNR is calculated using semblance and a 13-channel x 19-sample 2D moving window.}
    \label{fig:fig7}
\end{figure*}

From Figure \ref{fig:fig6}, it is clear that performance of all methods on out-of-sample data is similar to that on in-sample data (Fig \ref{fig:fig3}), with DAS-N2N unequivocally performing the greatest degree of noise suppression (Fig \ref{fig:fig6}E). This suggests that the DAS-N2N model has been adequately trained to generalize to sections of data outside of the training set and can be used for continual monitoring for this specific deployment. The DAS-N2N model also yields the highest local SNR for all three S-wave arrivals (again, regardless of window size or chosen summary statistic; Fig \ref{fig:fig7}E), with bandpass filtering also performing better than Wiener filtering and the jDAS model (Figs \ref{fig:fig7}B - D).

\begin{figure*}
    \centering
    \includegraphics[width=13.3cm]{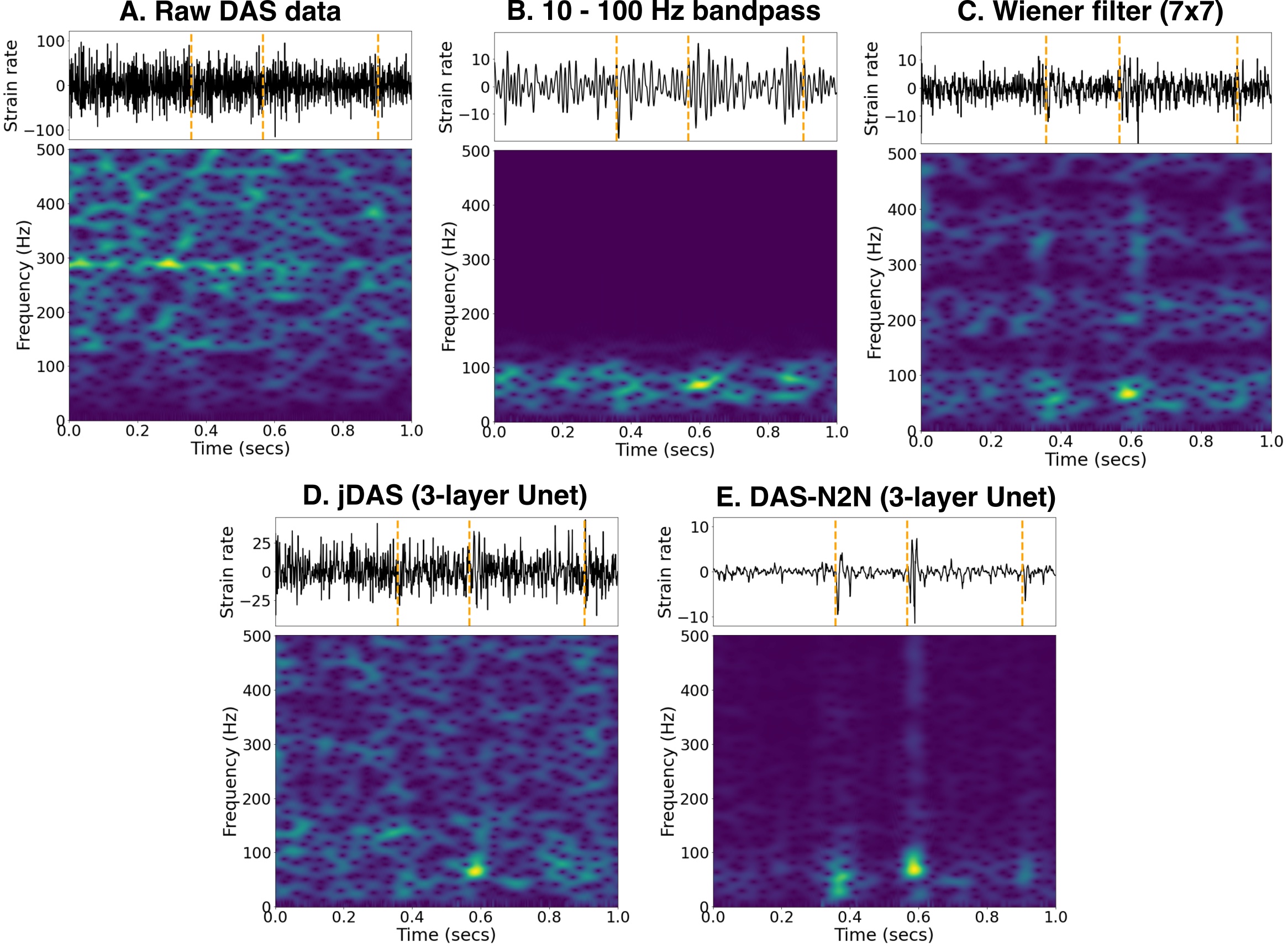}
    \caption{Individual DAS trace (top) and corresponding spectrogram (bottom) for DAS channel 587 in each example in Figure \ref{fig:fig6}. Strain rate is recorded in units of strain/s (counts).}
    \label{fig:fig8}
\end{figure*}

As with the in-sample data, Figure \ref{fig:fig8} shows the trace and spectrogram for an individual DAS channel processed by each method for this out-of-sample section of data. The three S-wave arrivals are difficult to discern in any of the traces or spectrograms except for in the DAS-N2N case (Fig \ref{fig:fig8}E), where all three arrivals appear as distinct features in both the time (top panel) and time-frequency (bottom panel) domains. Again, the raw observational noise appears to broadly follow a blue noise process, albeit with an apparent high frequency ‘ridge’ at approx. 285 Hz (Fig \ref{fig:fig8}A). Unlike the other methods, our DAS-N2N model adequately suppresses noise across the full spectrum (including the frequency band encompassing the phase arrivals) and retains weaker high-frequency signal components (Fig \ref{fig:fig8}E), a feat beyond the capability of standard spectral filtering methods. The approximate 4/5 scaling of absolute signal amplitudes (i.e., signal leakage) for DAS-N2N and Wiener filtering when compared with bandpass filtering is also present in this example.

\section{Discussion}
\label{sec:discussion}

\begin{figure*}
    \centering
    \includegraphics[width=13.3cm]{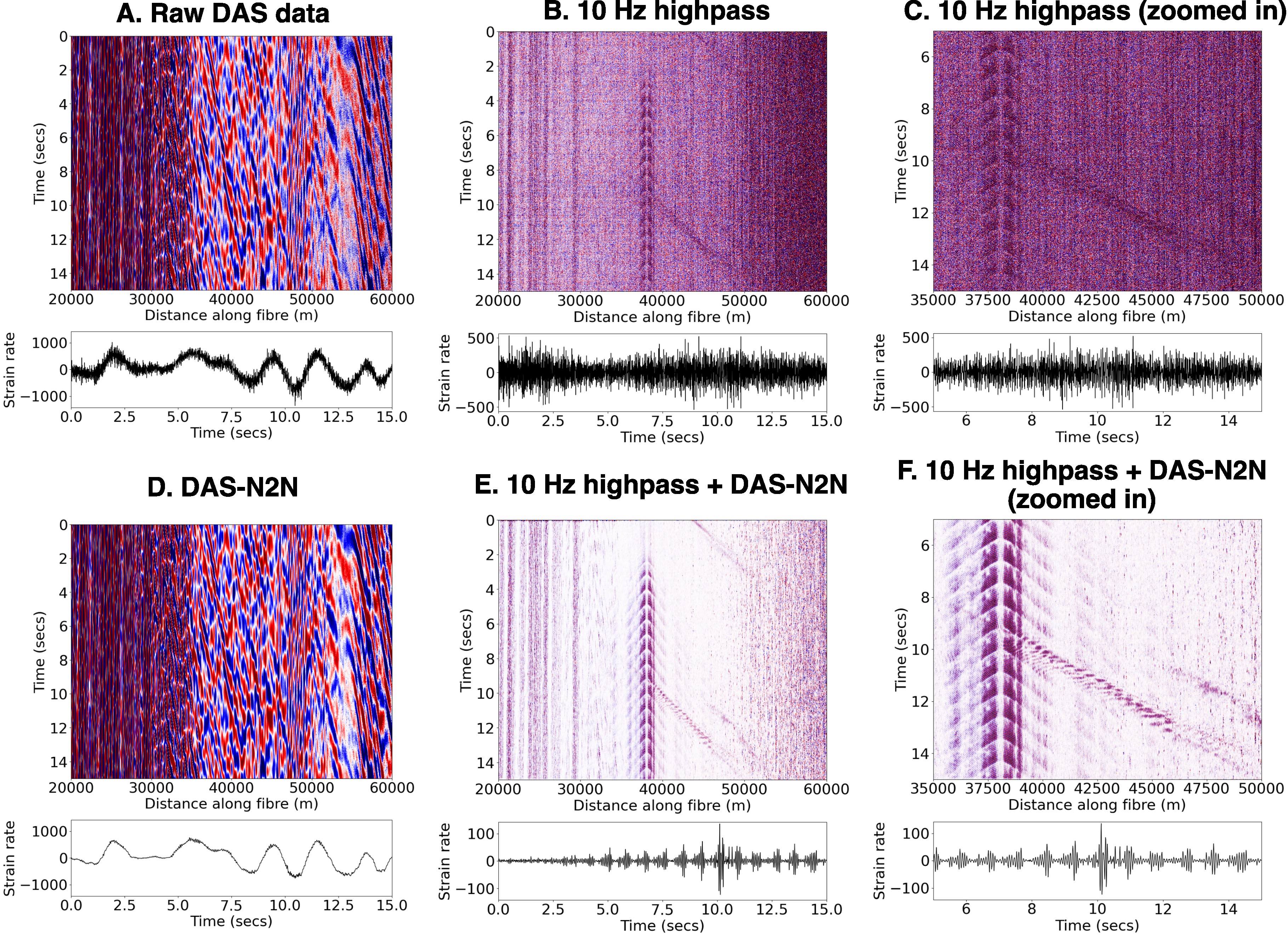}
    \caption{DAS data recorded by Wilcock \& OOI (2023) on submarine cable extending off the Oregon coast, USA (time given in seconds after 2021-11-02 10:36:09.839 UTC). Top: Data data for all DAS channels between 20km and 60km along south cable. Bottom: Data for individual DAS channel (40.06km along fibre). A) Raw DAS data. B) Butterworth (2-pass, 4th order) 10 Hz highpass filtered DAS data. C) Zoomed in version of highpass filtered data in B). D) DAS-N2N filtered DAS data. E) Application of Butterworth 10 Hz highpass filter, followed by DAS-N2N. F) Zoomed in version of highpass filtered + DAS-N2N processed data in E). Strain rate is recorded in units of strain/s (counts).}
    \label{fig:fig9}
\end{figure*}

In terms of random (incoherent) noise suppression, DAS-N2N unequivocally performs better than conventional Butterworth bandpass / Wiener filtering and a comparable self-supervised machine learning approach (jDAS) for the data presented here. This improved performance is immediately apparent in plots of the processed data (Figs \ref{fig:fig3} and \ref{fig:fig6}), where vertical bands of higher intensity noise over contiguous channels are suppressed only by DAS-N2N (as their locations differ between the two spliced fibres), and when estimates of SNR are determined through semblance (Figs \ref{fig:fig4} and \ref{fig:fig7}). Spectrograms from individual DAS channels (Figs \ref{fig:fig5} and \ref{fig:fig8}) show that part of this improved performance relates to the ability of machine learning models to suppress noise that lies in the same frequency band as the desired underlying signal. Such a feat will never be fully achievable for filters that rely on isolating or suppressing certain frequency bands, even when such techniques are enhanced through adaptive parameterization algorithms or the use of both temporal and spatial frequencies. Furthermore, common yet undesired causal filtering artefacts, such as precursory ringing before phase arrivals and signal polarity changes, will not be present in DAS-N2N processed data as the model processes the raw data directly and such features would only serve to increase model loss between the processed data and the target data. We note, however, that our DAS-N2N model does exhibit a degree of signal leakage, consistently reducing the absolute amplitude of the underlying signal by a factor of 1/5. Extensive experimentation with model depth, kernel size, choice of loss function (e.g., MAE, Huber), pre-processing steps (e.g., median removal and quantile normalization to reduce the impact of outliers), and architecture style (e.g., ResNet) did not yield any consistent improvement in this regard. As such, the issue of signal leakage is one that cannot be trivially solved here, and we leave this for future areas of research. It is worth mentioning that, regardless of this observed signal leakage, data processed by DAS-N2N exhibits higher signal-to-noise levels than any of the other methods presented, and its GPU-optimised implementation is also much more efficient (two of the primary factors controlling the effectiveness of subsequent imaging/event detection techniques and the viability of the method for processing large DAS datasets). Furthermore, once trained, our DAS-N2N model also shows an impressive degree of generalisation to other iDAS datasets, without the need for any retraining or fine-tuning (Fig \ref{fig:fig9}). 

Figure \ref{fig:fig9} shows application of our pre-trained Antarctica model on data collected during a 4-day DAS experiment conducted on two submarine cables extending off the US west coast from Pacific City, Oregon \citep{wilcock_ooi}. The south-most cable, which we examine here (Fig \ref{fig:fig9}), was interrogated by an iDASv3 DAS interrogator and extends over 80 km offshore. Large amplitude, long period ocean microseisms are clearly visible over background noise in both the unfiltered raw (Fig \ref{fig:fig9}A) and DAS-N2N processed (Fig \ref{fig:fig9}D) data. This is most apparent in the individual channel traces, where DAS-N2N filters strong high-frequency noise contaminating these long period signals (Figs \ref{fig:fig9}A and \ref{fig:fig9}D, bottom). Application of a 10 Hz highpass filter (Fig \ref{fig:fig9}B) reveals the presence of a lower amplitude blue whale 'A' call \citep[vertical pulse-like signal observed approx 40km along fibre;][]{wilcock_abadi_lipovsky_2023} and a much higher degree of incoherent noise as you go further along the fibre (due to decay of the interrogator light source). Subsequent application of DAS-N2N (Fig \ref{fig:fig9}E) greatly suppresses incoherent noise along the full extent of the fibre, revealing the individual pulses of the blue whale 'A' call (as well as other fin whale calls) in incredible detail (Fig \ref{fig:fig9}F). It is likely that DAS-N2N will also generalise well to other iDAS datasets (as this was the interrogator model used to acquire its training data), but will almost certainly need retraining to perform well on data collected by other interrogator models (due to differences in light source power, components used, measurement standards, etc).

By learning to map random noise to the distribution mean, DAS-N2N learns to perform the equivalent of a large-N stack (sum or average) over many noisy copies of the signal, analogous to the averaging of many short, independent, noisy exposures acquired during long-exposure low-light photography \citep{Lehtinen_Munkberg_Hasselgren_Laine_Karras_Aittala_Aila_2018}. The advantage of DAS-N2N over simple stacking, however, is that it only requires the acquisition of two noisy copies of the data for training, and only a single noisy copy of the data once trained. Furthermore, the noise in DAS-N2N processed data will be mapped to its distribution mean, whereas the noise in stacked data will only be mapped to its (statistically weaker) point-wise sample mean. 

The DAS-N2N approach, in general, is an order of magnitude faster than self-supervised ‘fill-in-the-gap’ approaches, such as jDAS (Figs \ref{fig:fig3} and \ref{fig:fig6}), as the latter’s masking procedure means it must process $N$ times more data (where $1/N$ is the fraction of input data masked). When compared with a jDAS model trained with the same model architecture, training hyperparameters and data pre-processing steps, DAS-N2N also performs better at the task of noise suppression on the microseismic icequake data presented here. In general, self-supervised learning methods will likely struggle to match or exceed the performance of weakly supervised learning methods, particularly on data with very high noise levels, as they are tasked with interpolating missing sections of data, which will always suffer from a degree of averaging over all possible values. On the other hand, weakly and fully supervised learning methods have the complete unmasked signal present in both the input and target data, meaning a direct 1-to-1 mapping can, theoretically, be learned.

In terms of computational efficiency, our 3-layer DAS-N2N model processes 30 s of recorded data in less than 1 s (Figs \ref{fig:fig3}E and \ref{fig:fig6}E) using the TensorFlow (version 2.3.0) Python framework and a single NVIDIA GeForce RTX 2080 Ti GPU. This is more than twice as fast as conventional channel-wise bandpass filtering using optimized low-level C programming language routines (Figs \ref{fig:fig3}B and \ref{fig:fig6}B). Any further algorithmic or filtering steps that yield improvements over bandpass filtering \citep{Isken_Vasyura-Bathke_Dahm_Heimann_2022, Chen_Savvaidis_Fomel_Chen_Saad_Wang_Oboue_Yang_Chen_2023} will obviously have further computational demands, making them increasingly less feasible for real-time passive monitoring purposes.

Arguably, the largest observed drawback of DAS-N2N against the other noise suppression methods presented is the degree of apparent signal leakage observed after data processing. This signal leakage is most likely a consequence of using an MSE loss function during training, but could also be due to unforeseen issues with our data pre- and post-processing steps (e.g., dividing and re-scaling by standard deviation of raw data) or an engineering aspect of the two fibres (e.g., channels on two fibres not lining up exactly). In any case, the degree of signal leakage appears to be consistent across observed signals in the data presented here and therefore, once a consistent scaling between DAS-N2N and bandpass filtered event signal amplitudes has been determined, one can apply a simple correction (e.g., for earthquake magnitude and source parameter estimation). However, we do not perform any correction here in order to keep processing methods as transparent, comparable and simple as possible. 

Another apparent drawback of all the methods presented is the inability to suppress unwanted coherent noise (e.g., the surface waves produced by the power generator for this DAS deployment). At present, this is likely still best performed by standard frequency filtering techniques (e.g., stop band / ‘notch’ filters), as such processes tend to produce signals with predictable and narrow-band frequency content \citep[e.g., 33 and 66 Hz for the power generator surface waves in Figs \ref{fig:fig3} and \ref{fig:fig6};][]{Hudson_Baird_Kendall_Kufner_Brisbourne_Smith_Butcher_Chalari_Clarke_2021}.

Finally, in terms of model architecture, we follow \citet{Lehtinen_Munkberg_Hasselgren_Laine_Karras_Aittala_Aila_2018} and \citet{van-den-Ende_Lior_Ampuero_Sladen_Ferrari_Richard_2021} in employing a simple U-Net architecture \citep{Ronneberger_Fischer_Brox_2015}. However, there are likely to be more effective model design choices for DAS-N2N and jDAS denoising than the ones chosen in these studies. Identifying optimal model architectures and training hyperparameters is often a challenging and sizeable task, involving either extensive manual trial-and-error or computationally expensive iterative search strategies \citep[e.g.,][]{Elsken_Metzen_Hutter_2019, Hutter_Kotthoff_Vanschoren_2019, White_Safari_Sukthanker_Ru_Elsken_Zela_Dey_Hutter_2023}. We therefore focus the scope of this article on the general applicability of N2N as a simple, effective strategy for denoising spliced-fibre DAS data without any clean training data or manual data curation. Furthermore, by demonstrating the effectiveness of DAS-N2N using a very small model (by deep learning standards), we provide evidence that DAS-N2N processing can be applied rapidly (well within ‘real-time’ constraints) and could be suitable for low-powered devices and edge networks.

\section{Conclusions}
\label{sec:conclusions}

In this article, we demonstrate the use of a weakly supervised machine learning method for fully automated random noise suppression in DAS data \citep[which we call DAS-N2N after the corresponding N2N technique in image processing;][]{Lehtinen_Munkberg_Hasselgren_Laine_Karras_Aittala_Aila_2018}. The method is ideally suited to DAS and other distributed optical fibre measurements (e.g., distributed temperature sensing; DTS) due to the ability to simultaneously record data across two spliced fibres within a single cable jacket. Advantageously, a DAS-N2N model can be trained end-to-end without any manual curation or labelling: simply, a section of data recorded on one of the spliced fibres serves as input data, with the corresponding section of data recorded on the other spliced fibre serving as target data (Fig \ref{fig:fig2}A and B). Once trained, the model only requires input data from a single un-spliced fibre (Fig \ref{fig:fig2}C), meaning there is no increase in data volumes to be stored after model training. Given the model's ability to generalise to other DAS settings, or if fibres can be temporarily (i.e., mechanically) or more permanently (i.e., fusion) spliced at some later point in time to facilitate model retraining, this approach can be applied retroactively to existing deployments with un-spliced fibres.

We demonstrate that DAS-N2N is inherently more effective and efficient than conventional bandpass filtering, Wiener filtering and self-supervised learning approaches. In particular, DAS-N2N is able to suppress noise lying within the same frequency range as the signal of interest (which is not possible for frequency-based filtering) and is an order of magnitude faster than self-supervised learning, due to the latter’s masking procedure. Furthermore, the presence of the complete unmasked underlying signal in both the input and target data when training a DAS-N2N model means that the signal can be retained through a 1-to-1 mapping, whereas self-supervised learning effectively performs a form of interpolation to predict the masked signal, which becomes more challenging as noise levels increase. Lastly, we demonstrate that a DAS-N2N model can be extremely lightweight (e.g., three model layers) and efficient, processing data in a fraction of the acquisition time (1/30 in the examples presented here) when optimized with a single GPU, and faster than standard frequency filtering routines optimized using compiled low-level programming languages, such as C. This offers the possibility of such models being further optimized, compiled and compressed for processing on low-powered devices and edge networks, which will be crucial for offshore or remote early warning monitoring settings.

\begin{acknowledgments}
    We thank NERC British Antarctic Survey for logistics and field support, with particular thanks to Sofia Kufner for her part in deploying the DAS in the field. We also thank Silixa for the loan of an iDAS interrogator. This work was funded by a NERC Collaborative Antarctic Science Scheme grant (grant number CASS-166), the BEAMISH project (grants: NE/G014159/1, NE/G013187/1), and the Digital Monitoring of CO2 storage project (DigiMon) (project no. 299622), which is part of the Accelerating CCS Technologies (ACT2) program. Author Sacha Lapins was funded by the DigiMon project and a Leverhulme Trust Early Career Fellowship. Authors Antony Butcher, Michael Kendall and Thomas Hudson were also funded by the DigiMon project. Author Maximilian Werner was funded by Natural Environment Research Council (NERC) grant NE/R017956/1 (“EQUIPT4RISK”). Author Jemma Gunning was supported by an EarthArt Fellowship at the School of Earth Sciences, University of Bristol. We are very grateful for comments and feedback provided by Feng Cheng and an anonymous reviewer. This work was carried out using the computational facilities of the Advanced Computing Research Centre, University of Bristol - http://www.bris.ac.uk/acrc/.
    
\end{acknowledgments}

\begin{dataavailability}
    The seismic data will be made available through the UK NERC Polar Data Centre. At the time of submission, the models, example data and code to reproduce the results in this paper were made available on GitHub (https://github.com/sachalapins/DAS-N2N), with the version associated with this paper archived through Zenodo \citep[][doi:10.5281/zenodo.7825684]{Lapins_Butcher_Kendall_Hudson_Stork_Brisbourne_2023}.
\end{dataavailability}

\bibliographystyle{gji}
\bibliography{dasn2n}

\newpage

\appendix
\section{Model architecture}

Table \ref{tab:tab1} gives a summary of the U-Net model architecture \citep{Ronneberger_Fischer_Brox_2015} used to implement DAS-N2N in this study. Prior to model training, model weights were initialized following \citet{pmlr-v9-glorot10a}. No batch normalization, dropout or other regularization techniques were used.

\begin{table*}
\begin{minipage}{100mm}
    \begin{tabular}{|l|c|c|l|}
         Layer name (type) & Output shape & Param \# & Function \\
         \hline
         input (InputLayer) & (128, 96, 1) & 0 &   \\
         conv00 (Conv2D) & (128, 96, 24) & 240 & Conv $3\times 3$ then LeakyReLU \\
         down10 (MaxPooling2D) & (64, 48, 24) & 0 & Max Pool $2\times 2$ \\
         conv10 (Conv2D) & (64, 48, 24) & 5208 & Conv $3\times 3$ then LeakyReLU \\
         up01 (UpSampling2D) & (128, 96, 24) & 0 & Upsample $2\times 2$ \\
         concat01 (Concatenate) & (128, 96, 48) & 0 & Concatenate with output of conv00 \\
         conv01a (Conv2D) & (128, 96, 48) & 20784 & Conv $3\times 3$ then LeakyReLU \\
         conv01b (Conv2D) & (128, 96, 48) & 20784 & Conv $3\times 3$ then LeakyReLU \\
         out01 (Conv2D) & (128, 96, 1) & 49 & Conv $1\times 1$ \\
         \hline
         \multicolumn{4}{|l|}{\textbf{Total trainable params:} 47,065}
    \end{tabular}
    \caption{Model architecture used to implement DAS-N2N in this study. Output shape given in rows $\times$ columns $\times$ feature maps. Param \# is number of trainable parameters in model layer. All convolutions use padding mode 'same', and except for the final layer are followed by leaky ReLU activation function with $\alpha = 0.1$. Final layer has linear activation. Upsample $2\times 2$ repeats data in each row and column.}
    \label{tab:tab1}
\end{minipage}
\end{table*}

\label{lastpage}

\end{document}